\documentclass[12pt,letterpaper]{article}
\pdfoutput=1
\usepackage{jheppub}
\usepackage{amsfonts, amsthm}
\usepackage[english]{babel}
\usepackage[utf8]{inputenc}
\usepackage{slashed}
\usepackage{mathrsfs}
\usepackage{amssymb}
\usepackage{color}
\hypersetup{unicode}
\usepackage{natbib}

\newcommand{\eq}{\begin{equation}}
\newcommand{\feq}{\end{equation}}
\newcommand{\eqn}{\begin{eqnarray}}
\newcommand{\feqn}{\end{eqnarray}}

\newcommand{\ma}[1]{\mbox{$\mathcal{#1}$}}

\newcommand{\mrm}[1]{\mbox{$\mathrm{#1}$}}
\newcommand{\D}{{\rm d}}

\title{AdS$_{\textbf{2}}$ solutions and their massive IIA origin}

\author[a]{Giuseppe Dibitetto}
\author[b]{Nicol\`o Petri}

\affiliation[a]{Institutionen f\"or fysik och astronomi, University of Uppsala,  Box 803, SE-751 08 Uppsala, Sweden.}
\affiliation[b]{Department of Mathematics, Bo\u{g}azi\c{c}i University,
34342, Bebek, Istanbul, Turkey.}

\emailAdd{giuseppe.dibitetto@physics.uu.se}
\emailAdd{nicolo.petri@boun.edu.tr}

\abstract{We consider warped $\mrm{AdS}_2\times \ma M_4$ backgrounds within $F(4)$ gauged supergravity in six dimensions. 
In particular, we are able to find supersymmetric solutions of the aforementioned type characterized by $\textrm{AdS}_{6}$ asymptotics and an $\ma M_4$ given by a three-sphere warped over a segment.
Subsequently, we provide the 10D uplift of the solutions to massive type IIA supergravity, where the geometry is $\mrm{AdS}_2\times S^{3}\times\tilde{S}^3$
warped over a strip. Finally we construct the brane intersection underlying one of these supergravity backgrounds. 
The explicit setup involves a D0-F1-D4 bound state intersecting a D4-D8 system.}

\begin{document}
\maketitle
\flushbottom

\section{Introduction}

Ever since the birth of the AdS/CFT correspondence \cite{Maldacena:1997re,Witten:1998qj}, the quest for supersymmetric AdS vacua in string theory has 
become a goal of utmost importance. All the research efforts in the last decades devoted to this task have delivered a wide range of results including
partial or exhaustive classifications of AdS string vacua in diverse dimensions 
(see \emph{e.g.} \cite{Grana:2004bg,Lust:2004ig,Gauntlett:2004zh,Gauntlett:2004hs,Bovy:2005qq,Grana:2006kf,Kounnas:2007dd,Tomasiello:2007eq,Koerber:2008rx,Caviezel:2008ik,Apruzzi:2013yva,Apruzzi:2014qva,Passias:2017yke,Passias:2018zlm,Cordova:2018eba}). 
A further crucial element for providing a holographic interpretation
of the corresponding AdS vacuum is to possess the underlying brane construction from which the solution emerges when taking the near-horizon limit (see \cite{Cvetic:2000cj} 
for a non-exhaustive collection of examples).

While in higher dimensions the organizing pattern of the landscape of supersymmetric AdS solutions is well delineated, achieving such a goal in two and
three dimensions turns out to be too hard of a task, at least in full generality. This is due to the vast and rich structure opening up when it comes 
to establishing the possible geometries and topologies of the internal space. However, some partial developments in this direction can be found in 
\cite{Argurio:2000tg,Kim:2005ez,Gauntlett:2006ns,Gauntlett:2006af,Donos:2008hd,Couzens:2017way,Eberhardt:2017uup}.
More recently in \cite{Dibitetto:2018gbk,Dibitetto:2018ftj} novel examples of $\textrm{AdS}_{2}$ \& $\textrm{AdS}_{3}$ solutions were found in the context 
of massive type IIA string theory. In all of the examples there the ten dimensional background is given by the warped product of AdS times a sphere 
warped over a line, their striking general feature being the non-compactness of the would-be internal space.

Such a feature seems to emerge very naturally within the context of brane intersections in massive type IIA supergravity when these produce $\textrm{AdS}_{2}$ or $\textrm{AdS}_{3}$ 
geometries in their near-horizon limit. So far a similar thing seems to be happening in higher dimensions only when the corresponding AdS vacua are obtained 
by employing non-Abelian T duality (NATD) as a generating technique (see \emph{e.g.} \cite{Itsios:2017cew,Lozano:2018pcp,Lozano:2018fvt} for recent examples of this type).
This issue makes the holographic interpretation of this type of supergravity constructions problematic. Nevertheless, within the context of \cite{Lozano:2018pcp},
the proposed holographic picture is that of an infinite quiver theory arising from a possible deconstructed extra dimension.

Going back to the original goal of finding novel examples of supersymmetric lower dimensional AdS vacua, a very fruitful approach seems to be that of 
exploiting the existence of consistent truncations of string and M theory yielding lower dimensional gauged supergravities as effective descriptions.
The reason why this can be so helpful is that one may restrict the search for solutions within a theory with a smaller amount of fields and excitations. 
Once in possession of new lower dimensional solutions, a ten (eleven) dimensional solution can be generated by using the needed uplift formula.
In this context, new supersymmetric solutions were found in \cite{Suh:2018tul,Hosseini:2018usu,Suh:2018szn} and \cite{Karndumri:2014hma,Rota:2015aoa,Apruzzi:2015zna,Dibitetto:2017klx,Dibitetto:2017tve,Karndumri:2018yiz}, 
by exploiting consistent truncations respectively down to $\mathcal{N}=(1,1)$ supergravity in six dimensions and $\mathcal{N}=1$ supergravity in seven dimensions.

The focus of this paper will be $F(4)$ supergravity in six dimensions as arising from a consistent truncation of massive type IIA supergravity on a 
squashed 4-sphere \cite{Cvetic:1999un}. We will study supersymmetric warped $\mrm{AdS}_2$ solutions supported both by a non-trivial 2-form field and a 
non-trivial profile for the universal scalar field. We will show how such an \emph{Ansatz} produces half-BPS solutions where the full six dimensional 
geometry is $\textrm{AdS}_{2}\times S^{3}$ warped over a segment. This, upon uplifting, will produce an $\mrm{AdS}_2$ solution in massive type IIA 
string theory where the ten dimensional background is given by $\textrm{AdS}_{2}\times S^{3}\times \tilde{S}^{3}$ warped over a strip.
Furthermore we will show how these ten dimensional backgrounds can be equally obtained by taking the near-horizon limit of a non-standard D0-F1-D4-D4$^\prime$-D8 
intersection specified by a certain brane charge distribution. Finally we will conclude by speculating on the physical interpretation of our construction.

\section{$\mrm{AdS}_2\times \ma M_4$ Solutions in $F(4)$ Gauged Supergravity}

In this section we derive two supersymmetric $\mrm{AdS}_2\times \ma M_4$ warped backgrounds in $d=6$ $F(4)$ gauged supergravity. The solutions preserve 8 real supercharges and, are characterized by an $\mrm{AdS}_6$ asymptotics and by a running profile for the 2-form gauge potential included in the supergravity multiplet. The 2-form wraps the internal directions of $\mrm{AdS}_2$ and supports the singular behaviors arising in the IR regime. As we will see this fact hints at the physical interpretation of these backgrounds in terms of branes intersecting the D4-D8 system giving rise to the $\mrm{AdS}_6$ vacuum.

We will firstly introduce our setup given by six dimensional $F(4)$ gauged supergravity in its minimal incarnation\footnote{We will restrict ourselves to the supergravity multiplet.}. Then we will formulate a suitable \emph{Ansatz} on the bosonic fields of the supergravity multiplet and on the corresponding Killing spinor.
With this information at hand, we will derive the BPS equations and we will solve them analytically.

\subsection{Minimal $\ma N=(1,1)$ Gauged Supergravity in $d=6$}

Minimal $\ma N=(1,1)$ supergravity in $d=6$ \cite{Romans:1985tw,Andrianopoli:2001rs} is obtained by only retaining the pure supergravity multiplet and,
as a consequence, the global isometry group breaks down to \cite{Andrianopoli:2001rs,Karndumri:2016ruc,Karndumri:2012vh}
\begin{equation}
 G_0=\mathbb R^+ \times \mathrm{SO}(4)\ .
 \label{global}
\end{equation}
The R-symmetry group is realized as the diagonal $\mathrm{SU}(2)_R \subset \mathrm{SO}(4)\simeq \mathrm{SU}(2)\times \mathrm{SU}(2)$. The corresponding $16$ supercharges of the theory are then organized in their irreducible chiral components. 
The fermionic field content of the supergravity multiplet is composed by two gravitini and two dilatini. Both the gravitini and the dilatini can be packed into pairs of Weyl spinors with opposite chiralities. Furthermore, in $d=1+5$ spacetime dimensions, symplectic-Majorana-Weyl spinors\footnote{For more details on Clifford algebras for $d=1+5$ spacetime dimensions see Appendix~\ref{app:SM_spinors}.} (SMW) may be introduced. The SMW formulation manifestly arranges the fermionic degrees of freedom of the theory into $\mathrm{SU}(2)_R$ doublets, which we respectively denote by $\psi_{\mu}^a$ and $\chi^a$ with $a=1,2$. It is worth mentioning that these objects have to respect a ``pseudo-reality condtion'' of the form in \eqref{SM_cond} in order for them to describe the correct number of propagating degrees of freedom.

The bosonic field content of the supergravity multiplet is given by the graviton $e_\mu^m$, a positive real scalar $X$, a 2-form gauge potential $\ma{B}_{(2)}$, a non-Abelian $\mathrm{SU}(2)$ valued vector field $A^i$ and an Abelian vector field $A^0$.
The consistent deformations of the minimal theory consist in a gauging of the R-symmetry $\mathrm{SU}(2)_R \subset \mathrm{SO}(4)$, by making use of the vectors $A^i$, and a St\"uckelberg coupling inducing a mass term for the 2-form field $\ma{B}_{(2)}$. The strength of the former deformation is controlled by a coupling constant $g$, while the latter by a mass parameter $m$.
The bosonic Lagrangian has the following form \cite{Romans:1985tw,Cvetic:1999un,Nunez:2001pt}
\begin{equation}
\begin{split}
\label{6dlagrangian}
 \ma L&=  R\,\star_{\,6}  1-4 \, X^{-2}\,\star_{\,6}\,dX\,\wedge \, dX-\frac12\,X^4\,\star_{\,6}\,\ma F_{(3)}\wedge \ma F_{(3)}-V(X)\\
 &- \frac12\, X^{-2} \,\left( \,\star_{\,6}\,\ma{F}_{(2)}^{i}\wedge \ma{F}_{(2)}^i+\star_{\,6}\,\ma{H}_{(2)}\wedge \ma{H}_{(2)}\,\right)-\frac12 \, \ma{B}_{(2)}\,\wedge \,\ma F_{(2)}^0\,\wedge \ma F_{(2)}^0\\
 &-\frac{1}{\sqrt 2}\,m \, \ma{B}_{(2)}\,\wedge \,\ma{B}_{(2)}\,\wedge \ma F_{(2)}^0-\frac13\,m^2 \, \ma{B}_{(2)}\,\wedge \,\ma{B}_{(2)}\,\wedge \ma{B}_{(2)}-\frac12 \, \ma{B}_{(2)}\,\wedge\, \ma{F}_{(2)}^{i}\wedge \ma{F}_{(2)}^i\,,
 \end{split}
\end{equation}
where the field strengths are defined as
\begin{equation}
 \begin{split}
  \mathcal{F}_{(3)} \ &= \ d\ma{B}_{(2)} \ ,\\
  \mathcal{F}_{(2)}^0 \ &= \ dA^0 \ ,\\
  \mathcal{H}_{(2)} \ &= \ dA^0+\sqrt{2}\,m\,\ma{B}_{(2)} \ ,\\
  \mathcal{F}_{(2)}^{i} \ &= \  d A^{i}  + \frac{g}{2} \, \epsilon^{ijk}\,A^{j}\wedge A^{k} \ .\\
 \end{split}
\end{equation}
A combination of the gauging and the massive deformation induces the following scalar potential 
\begin{equation}
 V(X)=m^2\,X^{-6}-4\sqrt{2}\,gm\,X^{-2}-2\,g^2\,X^2\ ,
 \label{scalarpotential}
\end{equation}
which can be re-expressed in terms of a real ``superpotential'' $f(X)$ through
\begin{equation}
 V(X)  = 16\,X^{2}\,\left(D_{X}f\right)^{2} -80\,f(X)^{2}\ ,
\end{equation}
where $f(X)$ is given by
\begin{equation}
 f(X)=\frac{1}{8}\,\left(m\,X^{-3}+\sqrt2 \,g\, X \right)\ .
 \label{superpotential}
\end{equation}

The supersymmetry variations of the fermionic fields are expressed in terms of a 6D Killing SMW spinor $\zeta^a$ as \cite{Romans:1985tw,Nunez:2001pt}
\begin{equation}
 \begin{split}
 \label{SUSYvariation6d}
  \delta_\zeta \psi_{\mu}^a\,=\,&\,\nabla_\mu\,\zeta^a+4g\,(A_\mu)^{a}_{\,\,\,b}\,\zeta^b+\frac{X^2}{48}\,\Gamma_*\Gamma^{mnp}\,\ma F_{(3)\,mnp}\,\Gamma_\mu\zeta^a\\
  &+i\,\frac{X^{-1}}{16\sqrt2}\,\left(\Gamma_\mu^{\,\,\,mn}-6\,e_\mu^m\,\Gamma^n\right)\,(\ma{\hat{H}}_{mn})^{a}_{\,\,\,b}\,\zeta^b-if(X)\,\Gamma_\mu\Gamma_*\zeta^a\ ,\\
  \delta_\zeta \chi^{a}\,=\,&\,X^{-1}\Gamma^m\partial_m X\,\zeta^a+\frac{X^2}{24}\,\Gamma_*\Gamma^{mnp}\,\ma F_{(3)\,mnp}\,\zeta^a\\
  &-i\,\frac{X^{-1}}{8\sqrt2}\,\Gamma^{mn}(\ma{\hat{H}}_{mn})^{a}_{\,\,\,b}\,\zeta^b+2i\,XD_X\,f(X)\,\Gamma_*\,\zeta^a\ ,
 \end{split}
\end{equation}
with $\nabla_\mu\,\zeta^a=\partial_\mu\zeta^a+\frac14\,\omega_\mu^{\,\,\,mn}\,\Gamma_{mn}\,\zeta^a$ and $(\ma{\hat{H}}_{mn})^{a}_{\,\,\,b}$ defined as
\begin{equation}
 (\ma{\hat{H}}_{\mu\nu})^{a}_{\,\,\,b}=\ma H_{(2)\,\mu\nu}\,\delta^{a}_{\,\,\,b}-4\,\Gamma_*\,(\ma F_{(2)\,\mu\nu})^{a}_{\,\,\,b}\ ,
\end{equation}
where we introduced the notation $A^{a}_{\,\,\,b}=\frac12\,A^i(\sigma^i)^{a}_{\,\,\,b}$, $\sigma^i$ being the Pauli matrices as given in \eqref{Pauli}.
By varying the Lagrangian \eqref{6dlagrangian} with respect to all the bosonic fields one obtains the following equations of motion
\begin{equation}
 \begin{split}
 &R_{\mu\nu}-4\,X^{-2}\,\partial_\mu X\,\partial_\nu\,X-\frac14\,V(X)\,g_{\mu\nu}-\frac14\,X^4\,\left(\ma F_{(3)\,\mu}^{\quad \,\,\, \alpha\beta}\ma F_{(3)\,\nu\alpha\beta}-\frac16\,\ma F_{(3)}^2\,g_{\mu\nu}\right)\\\vspace{2mm}
  &-\frac12\,X^{-2}\,\left(\ma H_{(2)\,\mu}^{\quad\,\,\,\alpha\,}\ma H_{(2)\,\nu\alpha}-\frac18\,\ma H_{(2)}^2g_{\mu\nu}\right)-\frac12\,X^{-2}\left(\ma F_{(2)\,\mu}^{i\quad\alpha}\,\ma F^i_{(2)\,\nu\alpha}-\frac18\ma F_{(2)}^{i\,\,\,2}\,g_{\mu\nu}\right)=0\,,\\\vspace{2mm}
  &d\left(X^4\,\star_{\,6}\,\ma F_{(3)}\right)=-\frac12\,\ma H_{(2)}\,\wedge\,\ma H_{(2)}-\frac12\,\ma F^i_{(2)}\,\wedge\,\ma F^i_{(2)}-\sqrt{2}\,m \,X^{-2}\,\star_{\,6}\,\ma H_{(2)}\,,\\\vspace{2mm}
  &d\left(X^{-2}\,\star_{\,6}\,\ma H_{(2)}\right)=-\ma H_{(2)}\,\wedge\,\ma F_{(3)}\,,\\\vspace{2mm}
  & D\left( X^{-2}\,\star_{\,6}\,\ma F_{(2)}^i\right)=-\ma F^i_{(2)}\,\wedge\,\ma F_{(3)}\,,\\\vspace{3mm}
 &d\left( X^{-1}\star_{\,6}dX\right)+\,\frac18\,X^{-2}\,\left(\star_{\,6}\,\ma H_{(2)}\,\wedge\,\ma H_{(2)}+\star_{\,6}\,\ma F^i_{(2)}\,\wedge\,\ma F^i_{(2)}\right)\\
 &-\frac14\,X^4\,\star_{\,6}\ma F_{(3)}\,\wedge\,\ma F_{(3)}-\frac18\,X\,D_X\,V(X)\star_{\,6}1=0\,,
 \end{split}
 \label{eom}
\end{equation}
where $D$ is the gauge covariant derivative defined as $D \omega^i=d\omega^i+g\,\epsilon_{ijk}\,A_{}^j\,\wedge\,\omega^k$ for any $\omega^i$ transforming covariantly with respect to $\mathrm{SU}(2)$.

Finally we mention that the scalar potential \eqref{scalarpotential} admits a critical point giving rise to an $\mathrm{AdS}_6$ vacuum preserving 16 real supercharges. This vacuum is realized by the following value of vev for $X$
\begin{equation}
 X=\frac{3^{1/4}\,m^{1/4}}{2^{1/8}\,g^{1/4}}\ ,
 \label{AdS6}
\end{equation}
while all the gauge potentials are zero.

\subsection{The General Ansatz}
\label{generalAnsatz}

Let us consider a 6D metric of the general form
\begin{equation}
 ds_6^2=e^{2U(\alpha)}\,ds^2_{{\scriptsize{\mrm{AdS}}_2}}+e^{2V(\alpha)}\,d\alpha^2+e^{2W(\alpha)}\,ds^2_{S^3}\ ,
 \label{general6dmetric}
\end{equation}
associated to a warped backgrounds of the type $\mrm{AdS}_2\times \ma M_4$ where $\ma M_4$ is locally written as a fibration of a $S^3$ over an interval $I_\alpha$. We point out that the warp factor $V$ is non-dynamical and it has been introduced because its gauge-fixing will turn out to be crucial to analytically solve the obtained BPS equations. 

 As far as the 2-form gauge potential $\ma{B}_{(2)}$ is concerned, it will purely wrap $\mrm{AdS}_2$ as follows
\begin{equation}
 \ma{B}_{(2)}=b(\alpha)\,\text{vol}_{{\scriptsize{\mrm{AdS}}_2}}\ .
 \label{general2form}
\end{equation}
We furthermore also assume a purely radial dependence for the scalar
\begin{equation}
 X=X(\alpha)\ ,
\end{equation}
and, for simplicity, we will restrict ourselves to the case of vanishing vectors, \emph{i.e.} $A^i=0$ and $A^0=0$.

We need also a suitable \emph{Ansatz} for the Killing spinor corresponding to the spacetime background given in \eqref{general6dmetric} and \eqref{general2form}. As we pointed out in \cite{Dibitetto:2018iar}, the action of the SUSY variations on the $\textrm{SU}(2)_R$ indices of the Killing spinor $\zeta^a$ is trivial, so it is more natural to cast the components of a Killing spinor in a $(1+5)$-dimensional Dirac spinor $\zeta$.
Following the splitting of the Clifford algebra given in \eqref{gammamatrices}, the Killing spinors considered are of the form
\begin{equation}
\begin{split}
 \zeta(\alpha)&=\zeta^+(\alpha)+\zeta^-(\alpha)\ ,\\
 \zeta^+&=i\,Y(\alpha)\left(\cos\theta(r)\,\chi^+_{{\scriptsize \mrm{AdS}_2}}\otimes \varepsilon_0+\,\sin\theta(r)\,\chi^+_{{\scriptsize \mrm{AdS}_2}}\,\otimes\sigma^3 \varepsilon_0  \right)\,\otimes\,\,\eta_{S^3}\ ,\\
 \zeta^-&=Y(\alpha)\left(\sin\theta(r)\,\chi^-_{{\scriptsize \mrm{AdS}_2}}\otimes \varepsilon_0-\,\cos\theta(r)\,\chi^-_{{\scriptsize \mrm{AdS}_2}}\,\otimes\sigma^3 \varepsilon_0  \right)\,\otimes\,\,\eta_{S^3}\ .
 \label{generalKilling}
 \end{split}
\end{equation}
 The spinor $\eta_{S^3}$ is a Dirac spinor, hence it has 4 real independent components and satisfies the following Killing equation
\begin{equation}
 \nabla_{\scriptsize{\theta^i}}\, \eta_{\,S^3}=\frac{i R}{2}\,\gamma_{\scriptsize{\theta^i}}\,\eta_{\,S^3}\ ,
 \label{killingM3}
\end{equation}
where $R^{-1}$ the radius of $S^3$ and $\gamma_{\scriptsize{\theta^i}}$ are the Dirac matrices introduced in \eqref{gammai} expressed in the curved basis $\{\theta^i\}$ on the 3-sphere.

Regarding the spinors $\chi^\pm_{{\scriptsize \mrm{AdS}_2}}$, they are Majorana-Weyl Killing spinors on $\mrm{AdS}_2$ and only possess 1 real independent component each. They respectively solve the equations\footnote{Since $\chi^\pm_{{\scriptsize \mrm{AdS}_2}}$ are Weyl spinor, they respectively satisfy the conditions $\Pi(\pm\rho_*)\chi^\pm_{{\scriptsize \mrm{AdS}_2}}=\pm\chi^\pm_{{\scriptsize \mrm{AdS}_2}}$ with $\Pi=\frac12(\mathbb{I}\pm \rho_*)$. It follows that they can organized in a Majorana doublet $\chi_{{\scriptsize \mrm{AdS}_2}}=(\chi^+_{{\scriptsize \mrm{AdS}_2}},\chi^-_{{\scriptsize \mrm{AdS}_2}})$ such that $
  \nabla_{\scriptsize{x^\alpha}}\, \chi_{{\scriptsize \mrm{AdS}_2}}=\frac{i L}{2}\,\rho_*\,\rho_{x^\alpha}\,\chi_{{\scriptsize \mrm{AdS}_2}}\ .\\ 
$}
\begin{equation}
 \nabla_{\scriptsize{x^\alpha}}\, \chi^\pm_{{\scriptsize \mrm{AdS}_2}}=\pm\,\frac{i L}{2}\,\rho_{x^\alpha}\,\chi^\mp_{{\scriptsize \mrm{AdS}_2}}\,,\\ 
 \label{killingAdS2}
\end{equation}
where $L^{-1}$ is the radius of $\mrm{AdS}_2$ and $\rho_{x^\alpha}$ are the Dirac matrices introduced in \eqref{rho3d} given in the curved basis $\{ x^\alpha \}$ on $\mathrm{AdS}_2$.

Finally $\varepsilon_0$ is a 2-dimensional real spinor encoding the two different chiral parts of $\zeta$ as
\begin{equation}
 \Gamma_*\,\zeta=\pm\,\zeta \qquad \Longleftrightarrow \qquad \sigma^3\,\varepsilon_0=\pm\,\varepsilon_0\ ,
\end{equation}
where we used the identity \eqref{gammastarrep}. Totally we have that $\zeta$ depends on 16 real independent supercharges that, as we will see, will be lowered by an algebraic projection condtion associated with the particular background considered.

\subsection{$\mathrm{AdS}_2\times S^3\times I_\alpha$ Warped Solutions}
\label{sol}

Let us now derive two analytic warped solutions of the type $\mathrm{AdS}_2\times S^3\times I_\alpha$ associated with the general background \ref{generalAnsatz}. Both preserve 8 real supercharges (BPS/2), enjoy an $\mrm{AdS}_6$ asymptotics and a singular IR regime.
The first solution is characterized by the following \emph{Ansatz},
\begin{equation}
 \begin{split}
   ds_6^2&=e^{2U(\alpha)}\,ds^2_{{\scriptsize{\mrm{AdS}}_2}}+e^{2V(\alpha)}\,d\alpha^2+e^{2W(\alpha)}\,ds^2_{S^3}\ ,\\
   \ma{B}_{(2)}&=b(\alpha)\,\text{vol}_{{\scriptsize{\mrm{AdS}}_2}}\ ,\\
  X&=X(\alpha)\ .
  \label{AdS2S3ansatz}
 \end{split}
\end{equation}
If we now impose the algebraic condition
\begin{equation}
\sigma^2 \varepsilon_0=\varepsilon_0\,,
\label{proj}
\end{equation}
on the spinor $\zeta$, written in \eqref{generalKilling}, we can specify the SUSY variations of fermions \eqref{SUSYvariation6d} for the background \eqref{AdS2S3ansatz}. In this way we obtain the following set of BPS equations,
\begin{equation}
 \begin{split}
  & U^\prime=\frac{1}{4}\,e^{V}\,\cos(2\theta)^{-1}\left((5+3\cos(4\theta))\,f+6 \sin{(2\theta)}^2\,X\,D_Xf+L\,e^{-U}\sin(2\theta)   \right)\,,\\
  &W^\prime=-\frac{1}{4}\,e^{V}\,\cos(2\theta)^{-1}\left((-9+\cos(4\theta))\,f+2 \sin{(2\theta)}^2\,X\,D_Xf  -L\,e^{-U}\sin(2\theta)  \right)\,,\\
  &b^\prime=-\frac{e^{V+2 U}}{X^2}\,\cos(2\theta)^{-1}\,\left(L\,e^{-U}+2 \sin (2 \theta )\,\left(f+3\,X\,D_Xf     \right)\right)\,,\\
  &\theta^\prime=-e^{V}  \,\sin (2 \theta )\,\left(f-X\,D_Xf     \right)\,,\\
  &Y^\prime=\frac{Y}{8}\,e^{V}\,\cos(2\theta)^{-1}\left((5+3\cos(4\theta))\,f+6 \sin{(2\theta)}^2\,X\,D_Xf+L\,e^{-U}\sin(2\theta)   \right)\,,\\
  &X^\prime=-\frac14\, e^{V}\, X\,\cos(2\theta)^{-1}\,\left(L\,e^{-U}\sin(2\theta)+2\,\sin(2\theta)^2\,f+ (7+\cos(4\theta))\,X\,D_Xf   \right)\,.\\
\label{floweqAdS2S3}
  \end{split}
\end{equation}
In addition to the first-order equations, one has to impose the two additional constraints
\begin{equation}
 \begin{split}
  b&\overset{!}{=}-\frac{e^{2 U}}{m}\,X\,\left(L\,e^{-U}+2\,\sin(2\theta)\,\left(f-\,X\,D_Xf     \right)\right)\,,\\
  R&=\frac12\,e^{-U+W}\,L\,\cos(2\theta)^{-1}+ e^{W}\, \tan (2 \theta )\,\left(3\,f+X\,D_Xf\right)\,.
  \label{constraintsAdS2S3}
 \end{split}
\end{equation}
If the superpotential $f$ is given by \eqref{superpotential}, it is easy to see that the constraints \eqref{constraintsAdS2S3} are satified.
Let us now make the following gauge choice on the non-dynamical warp factor $V$
\begin{equation}
 e^{V}=\left(\sin{(2\theta)}\,(X\,D_Xf-f)  \right)^{-1}\ .
\end{equation}
Then, the equations \eqref{floweqAdS2S3} can be integrated analytically for $\alpha\in [0, \frac{\pi}{4}]$ and the corresponding solution is given by
\begin{equation}
 \begin{split}
  &e^{2U}=\frac{2^{1/2}}{3^{1/2}m^{1/2}}\left(2\sqrt2\,g+3L\,(\cos(4\alpha)-1)\right)^{1/2}\,\sin(2\alpha)^{-1}\sin(4\alpha)^{-1}\,,\\
  &e^{2W}=\left(2\sqrt2\,g-6L\,\sin(2 \alpha)^2\right)^{1/2}\,\tan(2\alpha)^{-1}\sin(2\alpha)^{-1}\,,\\
  &e^{2V}=\frac{4\,3^{3/2}}{m^{1/2}}\,\cos(2\alpha)\tan(2\alpha)^{-2} \left(\sqrt2\, g-3L\,\sin(2\alpha)^2\right)^{-3/2}\,,\\
   &b=\frac{\sqrt2\,g-3L}{3m}\sin(2\alpha)^{-1}\, \cos (2 \alpha)^{-2}\,,\\
   &X=3^{1/4}\,m^{1/4}\,\cos(2\alpha)^{-1/2}\,\left(\sqrt2\,g-3L\,\sin(2 \alpha)^2\right)^{-1/4}\,,\\
     &Y=\left(2\sqrt2\,g-6L\,\sin(2 \alpha)^2\right)^{1/8}\,\sin(\alpha)^{-1/2}\cos(2\alpha)^{-1/4}\,,\\
        &\theta=-\alpha\,,
   \label{flowAdS2S3}
 \end{split}
\end{equation}
for $g>0$ and $m>0$.
From the constraints in \eqref{constraintsAdS2S3} we obtain the relation
\begin{equation}
 R=-3^{1/4}2^{-1/4}\,g\,m^{1/4}+3^{1/4}2^{-3/4}\,m^{1/4}\,L\,.
 \label{explicitconstraindAdS2S3}
\end{equation}
relating the radii $R$ and $L$ of the background to the gauging parameters $g$ and $m$.
The solution \eqref{flowAdS2S3} endowed with the constraint \eqref{explicitconstraindAdS2S3} satisfies the equations of motion of $F(4)$ gauged supergravity written in \eqref{eom}.
Finally, if we take the $\alpha\rightarrow 0$ limit, the solution \eqref{flowAdS2S3} is locally described by the $\mrm{AdS}_6$ vacuum \eqref{AdS6}, while for $\alpha \rightarrow \frac{\pi}{4}$, the background becomes singular.

The second solution is simpler and it can be found by setting the two warp factors $U$ and $W$ of \eqref{AdS2S3ansatz} equal. In this case, we produce a curved domain wall solution charged under the 2-form. The \emph{Ansatz} in this case has the following form
\begin{equation}
 \begin{split}
  & ds_6^2=e^{2U(\alpha)}\,\left( ds^2_{{\scriptsize{\mrm{AdS}}_2}}+ds^2_{S^3}\right)+e^{2V(\alpha)}\,d\alpha^2\ ,\\
  & \ma{B}_{(2)}=b(\alpha)\,\text{vol}_{{\scriptsize{\mrm{AdS}}_2}}\ ,\\
  &X=X(\alpha)\ .
  \label{CDWansatz}
 \end{split}
\end{equation}
With this prescription the Killing spinor \eqref{generalKilling} boils down to
\begin{equation}
\begin{split}
 &\zeta^+=Y(\alpha)\left(i\, \chi^+_{{\scriptsize \mrm{AdS}_2}}\otimes \varepsilon_0-\chi^-_{{\scriptsize \mrm{AdS}_2}}\,\otimes\sigma^3 \varepsilon_0  \right)\,\otimes\,\,\eta_{S^3}\ .
 \label{killingCDW}
 \end{split}
\end{equation}
Imposing again the algebraic condition \eqref{proj} on \eqref{killingCDW}, and plugging the \emph{Ansatz} \eqref{CDWansatz} into the SUSY variations of fermions \eqref{SUSYvariation6d}, we obtain the following set of BPS equations
\begin{equation}
 \begin{split}
  &U^\prime=-2\,e^{V}\,f\,,\qquad Y^\prime=-Y\,e^{V}\,f\ ,\\
  &b^\prime=-\frac{e^{U+V}\,L}{X^2}\,,\qquad X^\prime=2\,e^{V}\,X^2\,D_Xf\ ,
  \label{chargedDWeq}
 \end{split}
\end{equation}
that must be supplemented with the constraints
\begin{equation}
 b\overset{!}{=}-\frac{e^{U}\,X\,L}{m}\ ,\qquad \text{and} \qquad L=2R\ .
\end{equation}
Also in this case these constraints are satisfied if the superpotential has the form of \eqref{superpotential}. If we now make the gauge choice
\begin{equation}
 e^V=\left(2\,X^2\,D_Xf\right)^{-1}\ ,
\end{equation}
it is easy to see that the equations \eqref{chargedDWeq} are solved by the following expressions
\begin{equation}
 \begin{split}
  e^{2U}=&\ \frac{1}{2^{1/3}g^{2/3}} \left(\frac{\alpha}{\alpha^4-1}\right)^{2/3}\ ,\\
   e^{2V}=&\ \frac{8}{g^2}  \left(\frac{\alpha^2}{\alpha^4-1}\right)^{2}\ ,\\
   Y=&\ \frac{1}{2^{1/12}g^{1/6}} \left(\frac{\alpha}{\alpha^4-1}\right)^{1/6}\ ,\\
   b=&\, -\frac{3\,L}{2^{2/3}g^{4/3}}\frac{\alpha^{4/3}}{(\alpha^4-1)^{1/3}}\ ,\\
   X=&\ \alpha\ ,
   \label{chargedDWsol}
 \end{split}
\end{equation}
with $\alpha$ running between $0$ and $1$ if we choose $m$ and $g$ such that $m=\frac{\sqrt{2}\,g}{3}$. The solution \eqref{chargedDWsol} solves the equations of motion \eqref{chargedDWsol} and, in the $\alpha\rightarrow 1$ limit, it locally reproduces the $\mrm{AdS}_6$ vacuum \eqref{AdS6} with $m=\frac{\sqrt2\,g}{3}$, while, in $\alpha\rightarrow 0$, it manifests a singular behavior.

\section{The Massive IIA Origin}

We will now move to the 10D origin of these backgrounds in massive type IIA supergravity. We will start by discussing their uplifts by using the 
formula in \cite{Cvetic:1999un}. Later, for the simpler case, we will also provide a brane solution which will allow us reinterpret the charged domain
wall \eqref{chargedDWsol} as a particular background with polarized branes.

\subsection{Uplifts and $\mrm{AdS}_2\times S^3\times S^3\times I_\alpha\times I_\xi$ Backgrounds}
\label{cveticpope}

In this section we present the consistent truncation of massive IIA supergravity around the $\mrm{AdS}_6\times S^4$ warped vacuum \cite{Cvetic:1999un} and we discuss the uplifts of the $\mrm{AdS}_2\times \ma M_4$ solutions obtained in section \ref{sol}. 
If one choose the 6D gauge parameters as it follows
\begin{equation}
m=\frac{\sqrt{2}\,g}{3}\ ,
\label{m&guplift}
\end{equation}
 the 6D equations of motion \eqref{eom} can be obtained from the following truncation \emph{Ansatz} of the 10d background\footnote{We use the string frame, while in \cite{Cvetic:1999un} the truncation \emph{Ansatz} is given in the Einstein frame. See appendix \ref{app:massiveIIA}.} \cite{Cvetic:1999un}
\begin{equation}
 \begin{split}
  ds^2_{10}=s^{-1/3}\,X^{-1/2}\,\Delta^{1/2}\,\left[ds^2_6+2g^{-2}\,X^{2}\,ds_{4}^2\right]\ ,
  \label{truncationansatz}
 \end{split}
\end{equation}
where $\Delta=Xc^2+X^{-3}s^2$ and $ds_4^2$ is the metric of a squashed 4-sphere locally written as a fibration of a 3-sphere $\tilde{S}^3$ over a segment,
\begin{equation}
ds_{4}^2=d\xi^2+\frac14\,\Delta^{-1}\,X^{-3}\,c^{2}\,\sum_{i=1}^3\,\left(\theta^i-gA^i  \right)^2\ ,
 \label{4-sphere}
\end{equation}
with $c=\cos\xi$ and $s=\sin \xi$. The 3-sphere included in \eqref{4-sphere} is deformed and it is expressed as a $\textrm{SU}(2)$ bundle with connections $A^i$ and $\theta^i$ left-invariant 1-forms\footnote{They are defined by the relation $d\theta^i=-\frac12\,\varepsilon_{ijk}\,d\theta^j\,\wedge\,d\theta^k$.}.
The fluxes and the dilaton are given by \cite{Cvetic:1999un}
\begin{equation}
\label{10dfluxes}
 \begin{split}
  F_{(4)}&=-\frac{\sqrt 2}{6}\,g^{-3}\,s^{1/3}\,c^3\,\Delta^{-2}\,U\,d\xi\,\wedge\,\epsilon_{(3)}-\sqrt{2}\,g^{-3}\,s^{4/3}\,c^4\,\Delta^{-2}\,X^{-3}\,dX\,\wedge\,\epsilon_{(3)}\\
  &-\sqrt2 \,g^{-1}\,s^{1/3}\,c\,X^4\,\star_{\,6}\ma F_{(3)}\,\wedge\,d\xi-\frac{1}{\sqrt 2}\,s^{4/3}\,X^{-2}\,\star_{\,6}\ma H_{(2)}\\
  &+\frac{g^{-2}}{\sqrt 2}\,s^{1/3}\,c\,\ma F_{(2)}^i\,h^i\,\wedge\,d\xi-\frac{g^{-2}}{4\sqrt2}\,s^{4/3}\,c^2\,\Delta^{-1}\,X^{-3}\,\epsilon_{ijk}\,\ma F_{(2)}^i\,\wedge h^j\wedge\,h^k\ ,\\
  F_{(2)}&=\frac{s^{2/3}}{\sqrt 2}\,\ma H_{(2)}\ ,\qquad H_{(3)}=s^{2/3}\,\ma F_{(3)}+g^{-1}\,s^{-1/3}\,c\,\ma H_{(2)}\,\wedge\,d\xi\ ,\\
  e^{\Phi}&=s^{-5/6}\,\Delta^{1/4}\,X^{-5/4}\ ,\qquad F_{(0)}=m\ .
 \end{split}
\end{equation}
where $U=X^{-6}\,s^2-3X^2\,c^2+4\,X^{-2}\,c^2-6\,X^{-2}$ and $\epsilon_{(3)}=h^1\,\wedge\,h^2\,\wedge\,h^3$ with $h^i=\theta^i-gA^i$.
The $\mrm{AdS}_6\times S^4$ warped vacuum of massive IIA is naturally obtained by uplifting the 6D vacuum \eqref{AdS6}. In particular, for $X=1$ and vanishing gauge potentials, the manifold \eqref{4-sphere} becomes a round 4-sphere\footnote{As outlined in \cite{Brandhuber:1999np}, this is only the upper hemisphere of a 4-sphere with a boundary appearing for $\xi\rightarrow 0$.}. From \eqref{10dfluxes} it follows that the $\mrm{AdS}_6\times S^4$ vacuum is supported by the 4-flux $F_{(4)}$ that, together with the dilaton, has the following form
\begin{equation}
 F_{(4)}=\frac{5\sqrt{ 2}}{6}\,g^{-3}\,s^{1/3}\,c^3\,d\xi\,\wedge\,\epsilon_{(3)}\ ,\qquad e^{\Phi}=s^{-5/6}\ .
\end{equation}
These are exactly the flux and dilaton configurations corresponding to the near-horizon of the localized D4-D8 system of \cite{Brandhuber:1999np, Cvetic:1999un}.

The uplifts of the $\mrm{AdS}_2$ warped solutions obtained in section \ref{sol} can be easily derived by plugging the explicit form of the 6D backgrounds \eqref{flowAdS2S3} and \eqref{chargedDWsol} into the truncation formulas \eqref{truncationansatz} and \eqref{10dfluxes}. In both cases one obtains a 10D background $\mrm{AdS}_2\times S^3 \times \tilde{S}^3$ fibered over two intervals parametrized by the 6D coordinate $\alpha$ and by the internal coordinate $\xi$.

In particular we can write the corresponding 10D metric of the charged domain wall solution \eqref{chargedDWsol} as
\begin{equation}
 \begin{split}
  ds^2_{10}=s^{-1/3}\,X(\alpha)^{-1/2}\,\Delta^{1/2}\,\left[e^{2U(\alpha)}\left( \,ds^2_{{\scriptsize \mrm{AdS}_2}}+ds^2_{S^3}\right)+e^{2V(\alpha)}\,d\alpha^2+2g^{-2}\,X(\alpha)^{2}\,ds_{4}^2\right]\ ,
  \label{upliftCDW}
 \end{split}
\end{equation}
where $ds_{4}^2$ is given by \eqref{4-sphere} in the particular case of vanishing vectors $A^i=0$, i.e.
\begin{equation}
ds_4^2=d\xi^2+\frac14\,\Delta^{-1}\,X(r)^{-3}\,c^{2}\,ds_{\tilde{S}^3}^2\ .
 \label{intspaceCDW}
\end{equation}
The fluxes $F_{(4)}$, $F_{(2)}$ and $H_{(3)}$ can be easily derived from \eqref{10dfluxes} by setting also the abelian 6D vector $A^0=0$, i.e. $\ma H_{(2)}=\sqrt2\,m\,\ma B_{(2)}$.

\subsection{D4-D8 System and $\mrm{AdS}_6$ Vacua}
%
In order to provide the explicit brane picture producing the 10D background \eqref{upliftCDW} in its near-horizon limit, as a preliminary analysis, we review how the $\textrm{AdS}_{6}$ vacuum is obtained as the near-horizon limit of the
D4-D8 intersection.

The complete brane system realizing this mechanism is sketched in table~\ref{Table:D4D8}.
\begin{table}[h!]
\renewcommand{\arraystretch}{1}
\begin{center}
\scalebox{1}[1]{
\begin{tabular}{c||c|c c c c|c||c c c c}
branes & $t$ & $\rho$ & $\varphi^{1}$ & $\varphi^{2}$ & $\varphi^{3}$ & $z$ & $r$ & $\theta^{1}$ & $\theta^{2}$ & $\theta^{3}$\\
\hline \hline
D4 & $\times$ & $-$ & $-$ & $-$ & $-$ & $-$ & $\times$ & $\times$ & $\times$ & $\times$ \\
D8 & $\times$ & $\times$ & $\times$ & $\times$ & $\times$ & $-$ & $\times$ & $\times$ & $\times$ & $\times$  
\end{tabular}
}
\end{center}
\caption{{\it The brane picture underlying the 5d SCFT described by D4- and D8-branes. The above system is $\frac{1}{4}$ -- BPS.}} \label{Table:D4D8}
\end{table}
The corresponding string frame supergravity background reads 
\begin{align}
ds_{10}^{2} & = H_{\textrm{D}4}^{-1/2}H_{\textrm{D}8}^{-1/2}\left(-dt^{2}\right)+H_{\textrm{D}4}^{1/2}H_{\textrm{D}8}^{1/2}dz^{2}+H_{\textrm{D}4}^{-1/2}H_{\textrm{D}8}^{-1/2}\left(dr^{2}+r^{2}ds_{S^{3}}^{2}\right)+ \nonumber\\[1mm]
& \ \ +H_{\textrm{D}4}^{1/2}H_{\textrm{D}8}^{-1/2}\left(d\rho^{2}+\rho^{2}ds_{\tilde{S}^{3}}^{2}\right)\ , \\[2mm]
e^{\Phi} & = H_{\textrm{D}4}^{-1/4}H_{\textrm{D}8}^{-5/4} \ , \qquad C_{(5)} \ = \ \left(H_{\textrm{D}4}^{-1}-1\right)\, dt\wedge\textrm{vol}_{(4)} \ , \\[2mm]
C_{(9)} & =  \left(H_{\textrm{D}8}^{-1}-1\right)\, dt\wedge\textrm{vol}_{(4)}\wedge\tilde{\textrm{vol}}_{(4)} \ ,
\end{align}
where $\textrm{vol}_{(4)}$ \& $\tilde{\textrm{vol}}_{(4)}$ represent the volume forms on the $\mathbb{R}^{4}$ factors respectively spanned by $(r,\theta^{i})$ and $(\rho,\varphi^{i})$.
The functions $H_{\textrm{D}4}$ \& $H_{\textrm{D}8}$ specify a semilocalized D4-D8 intersection \cite{Brandhuber:1999np} and their explicit form is given by
\begin{equation}
\begin{array}{lclcccclclc}
H_{\textrm{D}8} & = & Q_{\textrm{D}8}\,z & , & & \textrm{and} & & H_{\textrm{D}4} & = & 1+Q_{\textrm{D}4}\,\left(\rho^{2}+\frac{4}{9}Q_{\textrm{D}8}z^{3}\right)^{-5/3} & .
\end{array}
\end{equation}
The above background yields a warped product of $\textrm{AdS}_{6}$ and a half $S^{4}$ in the limit where
\begin{equation}
\begin{array}{lcccclcccclc}
z\,\rightarrow\,0 & , & & \textrm{and} & & \rho\,\rightarrow\,0 & , & & \textrm{while} & & \frac{z^{3}}{\rho^{2}}\,\sim\,\textrm{finite} & .
\end{array}
\end{equation}
In what follows we will consider the intersection of the D4-D8 system with a D0-F1-D4$^\prime$ bound state. The presence of these new branes will break the isometry group of the $\textrm{AdS}_{6}\times S^4$ vacuum producing the $\textrm{AdS}_{2}$ foliation.

\subsection{The D0-F1-D4$^\prime$-D4-D8 Brane Intersection}
Given the above stringy picture, the complete brane system realizing the $\textrm{AdS}_{2}$ slicing of the 10D background is sketched in table~\ref{Table:D0F1D4D4D8}.
\begin{table}[h!]
\renewcommand{\arraystretch}{1}
\begin{center}
\scalebox{1}[1]{
\begin{tabular}{c||c|c c c c|c||c c c c}
branes & $t$ & $\rho$ & $\varphi^{1}$ & $\varphi^{2}$ & $\varphi^{3}$ & $z$ & $r$ & $\theta^{1}$ & $\theta^{2}$ & $\theta^{3}$\\
\hline \hline
D4 & $\times$ & $-$ & $-$ & $-$ & $-$ & $-$ & $\times$ & $\times$ & $\times$ & $\times$ \\
D8 & $\times$ & $\times$ & $\times$ & $\times$ & $\times$ & $-$ & $\times$ & $\times$ & $\times$ & $\times$  \\
\hline\hline
D0 & $\times$ & $-$ & $-$ & $-$ & $-$ & $-$ & $-$ & $-$ & $-$ & $-$ \\
F1 & $\times$ & $-$ & $-$ & $-$ & $-$ & $\times$ & $-$ & $-$ & $-$ & $-$ \\
D4$^{\prime}$ & $\times$ & $\times$ & $\times$ & $\times$ & $\times$ & $-$ & $-$ & $-$ & $-$ & $-$  
\end{tabular}
}
\end{center}
\caption{{\it The brane picture underlying the 1d SCFT described by D0 -- F1  -- D4$^{\prime}$ branes ending on a D4 -- D8 system. The above intersection is $\frac{1}{8}$ -- BPS.}} \label{Table:D0F1D4D4D8}
\end{table}
The corresponding supergravity background is that of a non-standard brane intersection in the spirit of \cite{Boonstra:1998yu}, since there is no
transverse direction which is common to all branes in the system.
The explicit profile of the massive IIA supergravity fields in the string frame reads 
\begin{align}
ds_{10}^{2} & = H_{\textrm{D}0}^{-1/2}H_{\textrm{F}1}^{-1}H_{\textrm{D}4}^{-1/2}H_{\textrm{D}4^{\prime}}^{-1/2}H_{\textrm{D}8}^{-1/2}\left(-dt^{2}\right)+H_{\textrm{D}0}^{1/2}H_{\textrm{F}1}^{-1}H_{\textrm{D}4}^{1/2}H_{\textrm{D}4^{\prime}}^{1/2}H_{\textrm{D}8}^{1/2}dz^{2} \\[1mm]
& \ \ +H_{\textrm{D}0}^{1/2}H_{\textrm{D}4}^{-1/2}H_{\textrm{D}4^{\prime}}^{1/2}H_{\textrm{D}8}^{-1/2}\left(dr^{2}+r^{2}ds_{S^{3}}^{2}\right)+H_{\textrm{D}0}^{1/2}H_{\textrm{D}4}^{1/2}H_{\textrm{D}4^{\prime}}^{-1/2}H_{\textrm{D}8}^{-1/2}\left(d\rho^{2}+\rho^{2}ds_{\tilde{S}^{3}}^{2}\right)\ ,\nonumber \\[2mm]
e^{\Phi} & =H_{\textrm{D}0}^{3/4}H_{\textrm{F}1}^{-1/2}H_{\textrm{D}4}^{-1/4}H_{\textrm{D}4^{\prime}}^{-1/4}H_{\textrm{D}8}^{-5/4} \ , \qquad B_{(2)} \ = \ \left(H_{\textrm{F}1}^{-1}-1\right)\, dt\wedge dz \ , \\[2mm]
C_{(5)} & =  H_{\textrm{D}4^{\prime}}\left(H_{\textrm{D}4}^{-1}-1\right)\, dt\wedge\textrm{vol}_{(4)} \ + \ H_{\textrm{D}4}\left(H_{\textrm{D}4^{\prime}}^{-1}-1\right)\, dt\wedge\tilde{\textrm{vol}}_{(4)} \ ,\\[2mm]
C_{(1)} & =  H_{\textrm{D}8}\left(H_{\textrm{D}0}^{-1}-1\right)\, dt \ , \quad C_{(9)}  =  \left(H_{\textrm{D}8}^{-1}-1\right)\, dt\wedge\textrm{vol}_{(4)}\wedge\tilde{\textrm{vol}}_{(4)} \ ,
\end{align}
where the warp factors appearing in the above metric read
\begin{equation}
\left\{
\begin{array}{lcl}
H_{\textrm{D}0} & = & H_{\textrm{F}1} \ = \ 1 \, + \, \frac{Q_{1}}{\rho^{2}} \, + \, \frac{Q_{2}}{r^{2}} \ , \\[1mm]
H_{\textrm{D}4} & = & 1 \, + \, \frac{Q_{1}}{\rho^{2}} \ , \\[1mm]
H_{\textrm{D}4^{\prime}} & = & 1 \, + \, \frac{Q_{2}}{r^{2}} \ , \\[1mm]
H_{\textrm{D}8} & = & 1 \, + \, Q_{3} \, z \ .
\end{array}\right.
\label{AdS2_branes}
\end{equation}
If we now take the limit $\rho\rightarrow 0$ while keeping $(z,r)$ finite, the metric becomes
\begin{equation}
ds_{10}^{2}  =  H_{\textrm{D}4^{\prime}}^{-1/2}H_{\textrm{D}8}^{-1/2}\left[Q_{1} \left(ds_{\textrm{AdS}_{2}}^{2}+ds_{S^{3}}^{2}\right)+
H_{\textrm{D}4^{\prime}}dr^{2}+H_{\textrm{D}4^{\prime}}H_{\textrm{D}8}dz^{2}+r^{2}H_{\textrm{D}4^{\prime}}ds_{\tilde{S}^{3}}^{2}\right] \ ,
\end{equation}
where $L_{\textrm{AdS}_2}=1/2$, which is $\textrm{AdS}_{2}\times S^{3} \times \tilde{S}^{3}$ warped over the $(z,r)$ coordinates.
By comparing \eqref{AdS2_branes} with \eqref{truncationansatz}, one finds an explicit mapping between the $(z,r)$ coordinates and the $(\alpha,\xi)$ coordinates appearing in the uplift formula.
In particular, by comparing the warp factors in front of the $\textrm{AdS}_{2}\times S^{3}$ block of the metric and the two expressions of the 10D dilaton,
one gets the following two algebraic relations
\begin{equation}
\left\{
 \begin{array}{lclc}
 Q_{1}\,H_{\textrm{D}4^{\prime}}^{-1/2}H_{\textrm{D}8}^{-1/2} & \overset{!}{=} & s^{-1/3}\Delta^{1/2}X^{-1/2}e^{2U} & , \\
 H_{\textrm{D}4^{\prime}}^{-1/4}H_{\textrm{D}8}^{-5/4} & \overset{!}{=} & s^{-5/6}\Delta^{1/4}X^{-5/4} & ,
 \end{array}\right.
\end{equation}
which, once combined with the matching condition for the $\tilde{S}^{3}$ block, give
\begin{equation}
 \left\{
 \begin{array}{lclc}
 r & = & \frac{2}{\sqrt{g}}Q_{1}^{-3/4}e^{3U/2}X^{-1/2}\,c & , \\
 z & = & Q_{3}^{-1} \left(Q_{1}^{-1/2}e^{U}X s^{2/3}-1\right) & .
 \end{array}\right.
\label{coord_change}
\end{equation}
The complete forms of the two 10D backgrounds match through the coordinate change in \eqref{coord_change}, upon further identifying $Q_{3}=m$, together
with the condition \eqref{m&guplift} relating the couplings $g$ \& $m$.

\label{sec:final}

\section*{Acknowledgements}

NP would like to thank I. Bena, N. Bobev, Y. Lozano, J. Montero and C. Nunez for enlightening discussions. NP would also like to thank
the members of the Department of Theoretical Physics at Uppsala University for their kind and friendly hospitality while some parts of this work were being prepared.
The work of NP is supported by T{\"U}B\.{I}TAK (Scientific and Technological Research Council of Turkey). The work of GD is supported by the Swedish Research Council (VR).

\appendix

\section{Massive IIA Supergravity}
\label{app:massiveIIA}

In this appendix we review the main features of massive IIA supergravity \cite{Romans:1985tz}. The theory is characterized by the bosonic fields $g_{MN}$, $\Phi$, $B_{(2)}$, $C_{(1)}$ and $C_{(3)}$. The action has the following form
\begin{equation}
  S_{\mathrm{mIIA}}=\frac{1}{2\kappa_{10}^2}\,\biggl [\int \D^{10}x\,\sqrt{-g}\,e^{-2\Phi}\left(R+4\,\partial_{\mu}\,\Phi\,\partial^{\mu}\,\Phi-\frac12\,|H_{(3)}|^2   \right)-\frac12 \sum_{p=0,2,4} |F_{(p)}|^2  \biggr]+S_{\text{top}}\ ,
  \label{massiveIIAaction}
\end{equation}
where $S_{\text{top}}$ is a topological term given by
\begin{equation}
\begin{split}
S_{\text{top}}=&-\frac{1}{2} \int ( B_{(2)}\wedge F_{(4)}\wedge F_{(4)}-\frac13  F_{(0)}\wedge B_{(2)}\wedge B_{(2)}\wedge B_{(2)}\wedge F_{(4)}\\
&+\frac{1}{20}F_{(0)}\wedge F_{(0)}\wedge B_{(2)}\wedge B_{(2)}\wedge B_{(2)}\wedge B_{(2)}\wedge B_{(2)})\ ,
\end{split}
\end{equation}
where $H_{(3)}=dB_{(2)}$, $F_{(2)}=dC_{(1)}$, $F_{(3)}=dC_{(3)}$ and the 0-form field strength $F_{(0)}$ is associated to the Romans' mass as $F_{(0)}=m$.

All the equations of motion can be derived\footnote{We set $\kappa_{10}=8\pi G_{10}=1$.} consistently from \eqref{massiveIIAaction}. They have the following form
\begin{equation}
 \begin{split}
  R_{MN}-\frac12\,T_{MN}&=0\ ,\\[1mm]
  \Box\Phi-|\partial \Phi|^2+\frac14R-\frac18 |H_{(3)}|^2&=0\ ,\\[1mm]
  d\left(e^{-2\Phi}\star_{10}H_{(3)}\right)&=0\ ,\\[1mm]
  \left(d+H_{(3)}\wedge\right)(\star_{10}F_{(p)})&=0\ ,\quad \text{with}\quad p=2,4\ ,
 \end{split}
 \label{massiveIIAeoms}
\end{equation}
where $M, N, \dots=0,\dots, 9$ and $R$ and $\Box$ are respectively the 10D scalar curvature and the Laplacian. The stress-energy tensor is given by
\begin{equation}
\begin{split}
 T_{MN}&=e^{2\Phi}\sum_p \left(\frac{p}{p!}\,F_{(p)\,MM_1\dots M_{p-1}}F_{(p)\,N}^{\qquad M_1\dots M_{p-1}}-\frac{p-1}{8}g_{MN}|F_{(p)}|^2 \right)\\
 &+\left(\frac12\,H_{(3)\,MPQ}H_{(3)\,N}^{\qquad PQ}-\frac14 g_{MN}|H_{(3)}|^2 \right)-\left(4\nabla_M \nabla_N \Phi +\frac12 g_{MN}(\Box \Phi-2|\partial \Phi|^2) \right)\,,
 \end{split}
\end{equation}
with $\nabla_M$ being associated with the Levi-Civita connection of the 10D background. The Bianchi identities take the form
\begin{equation}
\begin{split}
  dF_{(2)}&= F_{(0)}\wedge H_{(3)}\ ,\\
  d F_{(4)}&=-F_{(2)}\wedge H_{(3)}\ ,\\
 d H_{(3)}&=0\ ,\\
  d F_{(0)}&=0\ .
 \label{massiveIIAbianchi}
\end{split}
\end{equation}
As a consequence of \eqref{massiveIIAbianchi}, the following fluxes 
\begin{equation}
 F_{(0)}=m,\qquad H_{(3)}\,\qquad F_{(2)}-m B_{(2)}\,,\qquad F_{(4)}-B_{(2)}\wedge F_{(2)}+\frac12\,mB_{(2)}\wedge B_{(2)}\,,
\end{equation}
turn out to satisfy a Dirac quantization condition.

It may be worth mentioning that the truncation \emph{Ansatz} of section \ref{cveticpope} is obtained by casting massive IIA supergravity into the Einstein frame \cite{Cvetic:1999un}. To convert the action \eqref{massiveIIAaction}, the equations of motions \eqref{massiveIIAeoms} and Bianchi identities \eqref{massiveIIAbianchi} into the Einstein frame, one has to redefine the metric as $g_{MN}=e^{\Phi/2}\,g^{(\mathrm{E})}_{MN}$.

\section{Symplectic-Majorana-Weyl Spinors in $d=1+5$}
\label{app:SM_spinors}

In this appendix we collect the conventions and the fundamental relations involving irreducible spinors in $d=1+5$. Subsequently, we construct an explicit representation of Dirac matrices. In $d=1+5$ Dirac spinors enjoy 16 independent real components and they can be decomposed into irreducible Weyl spinors with opposite chirality and having 8 independent real components each. The 6D Clifford algebra is defined by the relation
\begin{equation}
\left\{\Gamma^{m},\,\Gamma^{n}\right\} = 2\,\eta^{mn} \, \mathbb{I}_{8} \ ,
\label{clifford6d}
\end{equation}
where $\left\{\Gamma^{m}\right\}_{m\,=\,0,\,\cdots\,5}$ are the $8\times 8$ Dirac matrices and $\eta=\text{diag}(-1,+1,+1,+1,+1)$. The chirality operator $\Gamma_*$ can be defined in the following way in terms of the above Dirac matrices 
\begin{equation}
\label{gammastar}
 \Gamma_*=\Gamma^0\,\Gamma^1\,\Gamma^2\,\Gamma^3\,\Gamma^4\,\Gamma^5\qquad \text{with}\qquad \Gamma_*\,\Gamma_*=\mathbb{I}_{8}\ .
\end{equation}
For $(1+5)$-dimensional backgrounds, we can choose the matrices $A, B, C$, respectively realizing Dirac, complex and charge conjugation, satisfying the following defining relations \cite{VanProeyen:1999ni}
\begin{equation}
 \left(\Gamma^{m}\right)^{\dagger}  =  -A \, \Gamma^{m} \, A^{-1} \,,\quad   \left(\Gamma^{m}\right)^{*}  = B \, \Gamma^{m} \, B^{-1}\,, \quad  \left(\Gamma^{m}\right)^{T}  =  -C \, \Gamma^{m} \, C^{-1} \ ,
\end{equation}
with 
\begin{equation}
 B^{T} = C \, A^{-1} \ ,\quad  B^{*}\,B  = -\mathbb{I}_{8} \ ,\quad C^{T}  =  -C^{-1}  =  -C^{\dagger}  =  C \  .
 \label{ABCidentities}
\end{equation}
The second identity in \eqref{ABCidentities} implies that it is actually inconsistent to define a proper reality condition on Dirac (or Weyl) spinors.
However, it is always possible to introduce $\mathrm{SU}(2)_R$ doublets $\zeta^a$ of Dirac spinors, called symplectic-Majorana (SM) spinors respecting a pseudo-reality condition \cite{VanProeyen:1999ni} given by
\begin{equation}
 \label{SM_cond}
\zeta_{a} \equiv\left(\zeta^{a}\right)^{*}  \overset{!}{=}  \epsilon_{ab}\,B\,\zeta^{b} \  ,
\end{equation}
where $\epsilon_{ab}$ is the $\mathrm{SU}(2)$ invariant Levi-Civita symbol. The condition \eqref{SM_cond} ensures us that the number of independent components of a SM spinor be the same of those of a Dirac spinor. Moreover, the above condition also turns out to be compatible with the projections onto the chiral components of a Dirac spinor. Hence it is possible to construct SM doublets of irreducible Weyl spinors that are called symplectic-Majorana-Weyl (SMW) spinors.

Let us now construct an explicit representation for the Dirac matrices satisfying \eqref{clifford6d}. We firstly introduce the Dirac matrices $\left\{\rho^\alpha \right\}_{\alpha\,=\,0,\,1}$ for a $(1+1)$-dimensional background in the following representation
\begin{equation}
 \rho^0=i\sigma^2\ ,\qquad  \rho^1=\sigma^1\ ,\qquad  \rho_*=\rho^0\rho^1=\sigma^3\ ,
 \label{rho3d}
\end{equation}
and the Dirac matrices for a Euclidean 3-dimensional background $\left\{\gamma^i \right\}_{i\,=1\,,2\,,3}$ as
\begin{equation}
 \gamma^1=\sigma^1\ ,\qquad \gamma^2=\sigma^2\ ,\qquad \gamma^3=\sigma^3\ ,
 \label{gammai}
\end{equation}
where
\begin{equation}
 \label{Pauli}
\sigma^{1} =
\left(
\begin{array}{cc}
0 & 1  \\
1 & 0
\end{array}
\right)
\hspace{5mm} \textrm{ , } \hspace{5mm}
\sigma^{2} =
\left(
\begin{array}{cc}
0 & -i  \\
i & 0
\end{array}
\right)
\hspace{5mm} \textrm{ , } \hspace{5mm}
\sigma^{3} =
\left(
\begin{array}{cc}
1 & 0  \\
0 & -1
\end{array}
\right) \ .
\end{equation}
are the Pauli matrices. An explicit representation of the $(1+5)$-dimensional Dirac matriced satisying \eqref{clifford6d} can be defined in the following way
\begin{equation}
 \begin{split}
  \Gamma^\alpha&=\rho^\alpha \,\otimes \, \sigma^1 \,\otimes \, \mathbb{I}_2 \ ,\qquad \text{with}\qquad \alpha=0,\,1\, ,\\
   \Gamma^2&=\rho_*\,\otimes \, \sigma^1\,\otimes \,\mathbb{I}_2\ ,\\
    \Gamma^i&=\mathbb{I}_2 \,\otimes \, \sigma^2 \,\otimes \, \gamma^i\ ,\qquad \,\text{with}\qquad \,i=1,\,2\,,3\, .
    \label{gammamatrices}
 \end{split}
\end{equation}
In this representation the chirality operator \eqref{gammastar} takes the form
\begin{equation}
 \Gamma_*=\mathbb{I}_2 \,\otimes \, \sigma^3 \,\otimes \, \mathbb{I}_2 \ ,
 \label{gammastarrep}
\end{equation}
while the matrices $A, B, C$ may be written as
\begin{equation}
 \begin{split}
  A&=\Gamma^0=i\,\sigma^2\,\,\otimes \, \sigma^1\otimes \, \mathbb{I}_2 \ ,\\ 
  B&=-i\,\Gamma^4\,\Gamma^5=\mathbb{I}_2 \,\otimes \, \mathbb{I}_2\,\otimes \, \sigma^1\  ,\\
  C&=i\,\Gamma^0\,\,\Gamma^4\,\Gamma^5=i\,\sigma^2 \,\otimes \, \sigma^1\,\otimes \, \sigma^1 \ .
  \label{abcrep}
 \end{split}
\end{equation}

 \bibliographystyle{utphys}
  \bibliography{references}

\providecommand{\href}[2]{#2}\begingroup\raggedright\begin{thebibliography}{10}

\bibitem{Maldacena:1997re}
J.~M. Maldacena, ``{The Large N limit of superconformal field theories and
  supergravity},'' \href{http://dx.doi.org/10.1023/A:1026654312961,
  10.4310/ATMP.1998.v2.n2.a1}{{\em Int. J. Theor. Phys.} {\bfseries 38} (1999)
  1113--1133}, \href{http://arxiv.org/abs/hep-th/9711200}{{\ttfamily
  arXiv:hep-th/9711200 [hep-th]}}.
[Adv. Theor. Math. Phys.2,231(1998)].

\bibitem{Witten:1998qj}
E.~Witten, ``{Anti-de Sitter space and holography},''
  \href{http://dx.doi.org/10.4310/ATMP.1998.v2.n2.a2}{{\em Adv. Theor. Math.
  Phys.} {\bfseries 2} (1998) 253--291},
\href{http://arxiv.org/abs/hep-th/9802150}{{\ttfamily arXiv:hep-th/9802150
  [hep-th]}}.

\bibitem{Grana:2004bg}
M.~Grana, R.~Minasian, M.~Petrini, and A.~Tomasiello, ``{Supersymmetric
  backgrounds from generalized Calabi-Yau manifolds},''
  \href{http://dx.doi.org/10.1088/1126-6708/2004/08/046}{{\em JHEP} {\bfseries
  08} (2004) 046},
\href{http://arxiv.org/abs/hep-th/0406137}{{\ttfamily arXiv:hep-th/0406137
  [hep-th]}}.

\bibitem{Lust:2004ig}
D.~Lust and D.~Tsimpis, ``{Supersymmetric AdS(4) compactifications of IIA
  supergravity},'' \href{http://dx.doi.org/10.1088/1126-6708/2005/02/027}{{\em
  JHEP} {\bfseries 02} (2005) 027},
\href{http://arxiv.org/abs/hep-th/0412250}{{\ttfamily arXiv:hep-th/0412250
  [hep-th]}}.

\bibitem{Gauntlett:2004zh}
J.~P. Gauntlett, D.~Martelli, J.~Sparks, and D.~Waldram, ``{Supersymmetric
  AdS(5) solutions of M theory},''
  \href{http://dx.doi.org/10.1088/0264-9381/21/18/005}{{\em Class. Quant.
  Grav.} {\bfseries 21} (2004) 4335--4366},
\href{http://arxiv.org/abs/hep-th/0402153}{{\ttfamily arXiv:hep-th/0402153
  [hep-th]}}.

\bibitem{Gauntlett:2004hs}
J.~P. Gauntlett, D.~Martelli, J.~Sparks, and D.~Waldram, ``{Supersymmetric AdS
  backgrounds in string and M-theory},'' {\em IRMA Lect. Math. Theor. Phys.}
  {\bfseries 8} (2005) 217--252,
\href{http://arxiv.org/abs/hep-th/0411194}{{\ttfamily arXiv:hep-th/0411194
  [hep-th]}}.

\bibitem{Bovy:2005qq}
J.~Bovy, D.~Lust, and D.~Tsimpis, ``{N = 1,2 supersymmetric vacua of IIA
  supergravity and SU(2) structures},''
  \href{http://dx.doi.org/10.1088/1126-6708/2005/08/056}{{\em JHEP} {\bfseries
  08} (2005) 056},
\href{http://arxiv.org/abs/hep-th/0506160}{{\ttfamily arXiv:hep-th/0506160
  [hep-th]}}.

\bibitem{Grana:2006kf}
M.~Grana, R.~Minasian, M.~Petrini, and A.~Tomasiello, ``{A Scan for new N=1
  vacua on twisted tori},''
  \href{http://dx.doi.org/10.1088/1126-6708/2007/05/031}{{\em JHEP} {\bfseries
  05} (2007) 031},
\href{http://arxiv.org/abs/hep-th/0609124}{{\ttfamily arXiv:hep-th/0609124
  [hep-th]}}.

\bibitem{Kounnas:2007dd}
C.~Kounnas, D.~Lust, P.~M. Petropoulos, and D.~Tsimpis, ``{AdS4 flux vacua in
  type II superstrings and their domain-wall solutions},''
  \href{http://dx.doi.org/10.1088/1126-6708/2007/09/051}{{\em JHEP} {\bfseries
  09} (2007) 051},
\href{http://arxiv.org/abs/0707.4270}{{\ttfamily arXiv:0707.4270 [hep-th]}}.

\bibitem{Tomasiello:2007eq}
A.~Tomasiello, ``{New string vacua from twistor spaces},''
  \href{http://dx.doi.org/10.1103/PhysRevD.78.046007}{{\em Phys. Rev.}
  {\bfseries D78} (2008) 046007},
\href{http://arxiv.org/abs/0712.1396}{{\ttfamily arXiv:0712.1396 [hep-th]}}.

\bibitem{Koerber:2008rx}
P.~Koerber, D.~Lust, and D.~Tsimpis, ``{Type IIA AdS(4) compactifications on
  cosets, interpolations and domain walls},''
  \href{http://dx.doi.org/10.1088/1126-6708/2008/07/017}{{\em JHEP} {\bfseries
  07} (2008) 017},
\href{http://arxiv.org/abs/0804.0614}{{\ttfamily arXiv:0804.0614 [hep-th]}}.

\bibitem{Caviezel:2008ik}
C.~Caviezel, P.~Koerber, S.~Kors, D.~Lust, D.~Tsimpis, and M.~Zagermann, ``{The
  Effective theory of type IIA AdS(4) compactifications on nilmanifolds and
  cosets},'' \href{http://dx.doi.org/10.1088/0264-9381/26/2/025014}{{\em Class.
  Quant. Grav.} {\bfseries 26} (2009) 025014},
\href{http://arxiv.org/abs/0806.3458}{{\ttfamily arXiv:0806.3458 [hep-th]}}.

\bibitem{Apruzzi:2013yva}
F.~Apruzzi, M.~Fazzi, D.~Rosa, and A.~Tomasiello, ``{All AdS$_7$ solutions of
  type II supergravity},''
  \href{http://dx.doi.org/10.1007/JHEP04(2014)064}{{\em JHEP} {\bfseries 04}
  (2014) 064},
\href{http://arxiv.org/abs/1309.2949}{{\ttfamily arXiv:1309.2949 [hep-th]}}.

\bibitem{Apruzzi:2014qva}
F.~Apruzzi, M.~Fazzi, A.~Passias, D.~Rosa, and A.~Tomasiello, ``{AdS$_{6}$
  solutions of type II supergravity},''
  \href{http://dx.doi.org/10.1007/JHEP11(2014)099,
  10.1007/JHEP05(2015)012}{{\em JHEP} {\bfseries 11} (2014) 099},
  \href{http://arxiv.org/abs/1406.0852}{{\ttfamily arXiv:1406.0852 [hep-th]}}.
[Erratum: JHEP05,012(2015)].

\bibitem{Passias:2017yke}
A.~Passias, G.~Solard, and A.~Tomasiello, ``{$ \mathcal{N}=2 $ supersymmetric
  AdS$_{4}$ solutions of type IIB supergravity},''
  \href{http://dx.doi.org/10.1007/JHEP04(2018)005}{{\em JHEP} {\bfseries 04}
  (2018) 005},
\href{http://arxiv.org/abs/1709.09669}{{\ttfamily arXiv:1709.09669 [hep-th]}}.

\bibitem{Passias:2018zlm}
A.~Passias, D.~Prins, and A.~Tomasiello, ``{A massive class of $\mathcal{N} =
  2$ AdS$_4$ IIA solutions},''
  \href{http://dx.doi.org/10.1007/JHEP10(2018)071}{{\em JHEP} {\bfseries 10}
  (2018) 071},
\href{http://arxiv.org/abs/1805.03661}{{\ttfamily arXiv:1805.03661 [hep-th]}}.

\bibitem{Cordova:2018eba}
C.~Cordova, G.~B. De~Luca, and A.~Tomasiello, ``{AdS$_8$ Solutions in Type II
  Supergravity},''
\href{http://arxiv.org/abs/1811.06987}{{\ttfamily arXiv:1811.06987 [hep-th]}}.

\bibitem{Cvetic:2000cj}
M.~Cvetic, H.~Lu, C.~N. Pope, and J.~F. Vazquez-Poritz, ``{AdS in warped
  space-times},'' \href{http://dx.doi.org/10.1103/PhysRevD.62.122003}{{\em
  Phys. Rev.} {\bfseries D62} (2000) 122003},
\href{http://arxiv.org/abs/hep-th/0005246}{{\ttfamily arXiv:hep-th/0005246
  [hep-th]}}.

\bibitem{Argurio:2000tg}
R.~Argurio, A.~Giveon, and A.~Shomer, ``{Superstring theory on AdS(3) x G / H
  and boundary N=3 superconformal symmetry},''
  \href{http://dx.doi.org/10.1088/1126-6708/2000/04/010}{{\em JHEP} {\bfseries
  04} (2000) 010},
\href{http://arxiv.org/abs/hep-th/0002104}{{\ttfamily arXiv:hep-th/0002104
  [hep-th]}}.

\bibitem{Kim:2005ez}
N.~Kim, ``{AdS(3) solutions of IIB supergravity from D3-branes},''
  \href{http://dx.doi.org/10.1088/1126-6708/2006/01/094}{{\em JHEP} {\bfseries
  01} (2006) 094},
\href{http://arxiv.org/abs/hep-th/0511029}{{\ttfamily arXiv:hep-th/0511029
  [hep-th]}}.

\bibitem{Gauntlett:2006ns}
J.~P. Gauntlett, N.~Kim, and D.~Waldram, ``{Supersymmetric AdS(3), AdS(2) and
  Bubble Solutions},''
  \href{http://dx.doi.org/10.1088/1126-6708/2007/04/005}{{\em JHEP} {\bfseries
  04} (2007) 005},
\href{http://arxiv.org/abs/hep-th/0612253}{{\ttfamily arXiv:hep-th/0612253
  [hep-th]}}.

\bibitem{Gauntlett:2006af}
J.~P. Gauntlett, O.~A.~P. Mac~Conamhna, T.~Mateos, and D.~Waldram,
  ``{Supersymmetric AdS(3) solutions of type IIB supergravity},''
  \href{http://dx.doi.org/10.1103/PhysRevLett.97.171601}{{\em Phys. Rev. Lett.}
  {\bfseries 97} (2006) 171601},
\href{http://arxiv.org/abs/hep-th/0606221}{{\ttfamily arXiv:hep-th/0606221
  [hep-th]}}.

\bibitem{Donos:2008hd}
A.~Donos, J.~P. Gauntlett, and J.~Sparks, ``{AdS(3) x (S**3 x S**3 x S**1)
  Solutions of Type IIB String Theory},''
  \href{http://dx.doi.org/10.1088/0264-9381/26/6/065009}{{\em Class. Quant.
  Grav.} {\bfseries 26} (2009) 065009},
\href{http://arxiv.org/abs/0810.1379}{{\ttfamily arXiv:0810.1379 [hep-th]}}.

\bibitem{Couzens:2017way}
C.~Couzens, C.~Lawrie, D.~Martelli, S.~Schafer-Nameki, and J.-M. Wong,
  ``{F-theory and AdS$_{3}$/CFT$_{2}$},''
  \href{http://dx.doi.org/10.1007/JHEP08(2017)043}{{\em JHEP} {\bfseries 08}
  (2017) 043},
\href{http://arxiv.org/abs/1705.04679}{{\ttfamily arXiv:1705.04679 [hep-th]}}.

\bibitem{Eberhardt:2017uup}
L.~Eberhardt, ``{Supersymmetric AdS$_{3}$ supergravity backgrounds and
  holography},'' \href{http://dx.doi.org/10.1007/JHEP02(2018)087}{{\em JHEP}
  {\bfseries 02} (2018) 087},
\href{http://arxiv.org/abs/1710.09826}{{\ttfamily arXiv:1710.09826 [hep-th]}}.

\bibitem{Dibitetto:2018gbk}
G.~Dibitetto and A.~Passias, ``{AdS$_{2}\times$ S$^{7}$ solutions from D0-F1-D8
  intersections},'' \href{http://dx.doi.org/10.1007/JHEP10(2018)190}{{\em JHEP}
  {\bfseries 10} (2018) 190},
\href{http://arxiv.org/abs/1807.00555}{{\ttfamily arXiv:1807.00555 [hep-th]}}.

\bibitem{Dibitetto:2018ftj}
G.~Dibitetto, G.~Lo~Monaco, A.~Passias, N.~Petri, and A.~Tomasiello, ``{AdS$_3$
  solutions with exceptional supersymmetry},''
  \href{http://dx.doi.org/10.1002/prop.201800060}{{\em Fortsch. Phys.}
  {\bfseries 66} no.~10, (2018) 1800060},
\href{http://arxiv.org/abs/1807.06602}{{\ttfamily arXiv:1807.06602 [hep-th]}}.

\bibitem{Itsios:2017cew}
G.~Itsios, Y.~Lozano, J.~Montero, and C.~Nunez, ``{The AdS$_{5}$ non-Abelian
  T-dual of Klebanov-Witten as a $ \mathcal{N}=1 $ linear quiver from
  M5-branes},'' \href{http://dx.doi.org/10.1007/JHEP09(2017)038}{{\em JHEP}
  {\bfseries 09} (2017) 038},
\href{http://arxiv.org/abs/1705.09661}{{\ttfamily arXiv:1705.09661 [hep-th]}}.

\bibitem{Lozano:2018pcp}
Y.~Lozano, N.~T. Macpherson, and J.~Montero, ``{$AdS_6$ T-duals and Type IIB
  $AdS_6\times S^2$ Geometries with 7-Branes},''
\href{http://arxiv.org/abs/1810.08093}{{\ttfamily arXiv:1810.08093 [hep-th]}}.

\bibitem{Lozano:2018fvt}
Y.~Lozano, N.~T. Macpherson, and J.~Montero, ``{AdS$_6$ T-duals and AdS$_6
  \times S^2$ geometries in Type IIB},'' in {\em {Corfu2018: 18th Hellenic
  School and Workshops on Elementary Particle Physics and Gravity (Corfu2018)
  Corfu, Corfu, Greece, August 31-September 28, 2018}}.
\newblock 2018.
\newblock
\href{http://arxiv.org/abs/1811.00054}{{\ttfamily arXiv:1811.00054 [hep-th]}}.
\newblock

\bibitem{Suh:2018tul}
M.~Suh, ``{D4-branes wrapped on supersymmetric four-cycles},''
\href{http://arxiv.org/abs/1809.03517}{{\ttfamily arXiv:1809.03517 [hep-th]}}.

\bibitem{Hosseini:2018usu}
S.~M. Hosseini, K.~Hristov, A.~Passias, and A.~Zaffaroni, ``{6D attractors and
  black hole microstates},''
\href{http://arxiv.org/abs/1809.10685}{{\ttfamily arXiv:1809.10685 [hep-th]}}.

\bibitem{Suh:2018szn}
M.~Suh, ``{D4-branes wrapped on supersymmetric four-cycles from matter coupled
  $F(4)$ gauged supergravity},''
\href{http://arxiv.org/abs/1810.00675}{{\ttfamily arXiv:1810.00675 [hep-th]}}.

\bibitem{Karndumri:2014hma}
P.~Karndumri, ``{RG flows in 6D N = (1,0) SCFT from SO(4) half-maximal 7D
  gauged supergravity},'' \href{http://dx.doi.org/10.1007/JHEP06(2014)101}{{\em
  JHEP} {\bfseries 06} (2014) 101},
\href{http://arxiv.org/abs/1404.0183}{{\ttfamily arXiv:1404.0183 [hep-th]}}.

\bibitem{Rota:2015aoa}
A.~Rota and A.~Tomasiello, ``{AdS$_{4}$ compactifications of AdS$_{7}$
  solutions in type II supergravity},''
  \href{http://dx.doi.org/10.1007/JHEP07(2015)076}{{\em JHEP} {\bfseries 07}
  (2015) 076},
\href{http://arxiv.org/abs/1502.06622}{{\ttfamily arXiv:1502.06622 [hep-th]}}.

\bibitem{Apruzzi:2015zna}
F.~Apruzzi, M.~Fazzi, A.~Passias, and A.~Tomasiello, ``{Supersymmetric
  AdS$_{5}$ solutions of massive IIA supergravity},''
  \href{http://dx.doi.org/10.1007/JHEP06(2015)195}{{\em JHEP} {\bfseries 06}
  (2015) 195},
\href{http://arxiv.org/abs/1502.06620}{{\ttfamily arXiv:1502.06620 [hep-th]}}.

\bibitem{Dibitetto:2017klx}
G.~Dibitetto and N.~Petri, ``{6d surface defects from massive type IIA},''
  \href{http://dx.doi.org/10.1007/JHEP01(2018)039}{{\em JHEP} {\bfseries 01}
  (2018) 039},
\href{http://arxiv.org/abs/1707.06154}{{\ttfamily arXiv:1707.06154 [hep-th]}}.

\bibitem{Dibitetto:2017tve}
G.~Dibitetto and N.~Petri, ``{BPS objects in D = 7 supergravity and their
  M-theory origin},'' \href{http://dx.doi.org/10.1007/JHEP12(2017)041}{{\em
  JHEP} {\bfseries 12} (2017) 041},
\href{http://arxiv.org/abs/1707.06152}{{\ttfamily arXiv:1707.06152 [hep-th]}}.

\bibitem{Karndumri:2018yiz}
P.~Karndumri and P.~Nuchino, ``{Supersymmetric solutions of matter-coupled 7D
  N=2 gauged supergravity},''
  \href{http://dx.doi.org/10.1103/PhysRevD.98.086012}{{\em Phys. Rev.}
  {\bfseries D98} no.~8, (2018) 086012},
\href{http://arxiv.org/abs/1806.04064}{{\ttfamily arXiv:1806.04064 [hep-th]}}.

\bibitem{Cvetic:1999un}
M.~Cvetic, H.~Lu, and C.~N. Pope, ``{Gauged six-dimensional supergravity from
  massive type IIA},''
  \href{http://dx.doi.org/10.1103/PhysRevLett.83.5226}{{\em Phys. Rev. Lett.}
  {\bfseries 83} (1999) 5226--5229},
\href{http://arxiv.org/abs/hep-th/9906221}{{\ttfamily arXiv:hep-th/9906221
  [hep-th]}}.

\bibitem{Romans:1985tw}
L.~J. Romans, ``{The F(4) Gauged Supergravity in Six-dimensions},''
\href{http://dx.doi.org/10.1016/0550-3213(86)90517-1}{{\em Nucl. Phys.}
  {\bfseries B269} (1986) 691}.

\bibitem{Andrianopoli:2001rs}
L.~Andrianopoli, R.~D'Auria, and S.~Vaula, ``{Matter coupled F(4) gauged
  supergravity Lagrangian},''
  \href{http://dx.doi.org/10.1088/1126-6708/2001/05/065}{{\em JHEP} {\bfseries
  05} (2001) 065},
\href{http://arxiv.org/abs/hep-th/0104155}{{\ttfamily arXiv:hep-th/0104155
  [hep-th]}}.

\bibitem{Karndumri:2016ruc}
P.~Karndumri and J.~Louis, ``{Supersymmetric $AdS_6$ vacua in six-dimensional
  $N=(1,1)$ gauged supergravity},''
  \href{http://dx.doi.org/10.1007/JHEP01(2017)069}{{\em JHEP} {\bfseries 01}
  (2017) 069},
\href{http://arxiv.org/abs/1612.00301}{{\ttfamily arXiv:1612.00301 [hep-th]}}.

\bibitem{Karndumri:2012vh}
P.~Karndumri, ``{Holographic RG flows in six dimensional F(4) gauged
  supergravity},'' \href{http://dx.doi.org/10.1007/JHEP01(2013)134,
  10.1007/JHEP06(2015)165}{{\em JHEP} {\bfseries 01} (2013) 134},
\href{http://arxiv.org/abs/1210.8064}{{\ttfamily arXiv:1210.8064 [hep-th]}}.

\bibitem{Nunez:2001pt}
C.~Nunez, I.~Y. Park, M.~Schvellinger, and T.~A. Tran, ``{Supergravity duals of
  gauge theories from F(4) gauged supergravity in six-dimensions},''
  \href{http://dx.doi.org/10.1088/1126-6708/2001/04/025}{{\em JHEP} {\bfseries
  04} (2001) 025},
\href{http://arxiv.org/abs/hep-th/0103080}{{\ttfamily arXiv:hep-th/0103080
  [hep-th]}}.

\bibitem{Dibitetto:2018iar}
G.~Dibitetto and N.~Petri, ``{Surface defects in the D4 $-$ D8 brane system},''
\href{http://arxiv.org/abs/1807.07768}{{\ttfamily arXiv:1807.07768 [hep-th]}}.

\bibitem{Brandhuber:1999np}
A.~Brandhuber and Y.~Oz, ``{The D-4 - D-8 brane system and five-dimensional
  fixed points},'' \href{http://dx.doi.org/10.1016/S0370-2693(99)00763-7}{{\em
  Phys. Lett.} {\bfseries B460} (1999) 307--312},
\href{http://arxiv.org/abs/hep-th/9905148}{{\ttfamily arXiv:hep-th/9905148
  [hep-th]}}.

\bibitem{Boonstra:1998yu}
H.~J. Boonstra, B.~Peeters, and K.~Skenderis, ``{Brane intersections, anti-de
  Sitter space-times and dual superconformal theories},''
  \href{http://dx.doi.org/10.1016/S0550-3213(98)00512-4}{{\em Nucl. Phys.}
  {\bfseries B533} (1998) 127--162},
\href{http://arxiv.org/abs/hep-th/9803231}{{\ttfamily arXiv:hep-th/9803231
  [hep-th]}}.

\bibitem{Romans:1985tz}
L.~J. Romans, ``{Massive N=2a Supergravity in Ten-Dimensions},''
\href{http://dx.doi.org/10.1016/0370-2693(86)90375-8}{{\em Phys. Lett.}
  {\bfseries 169B} (1986) 374}.

\bibitem{VanProeyen:1999ni}
A.~Van~Proeyen, ``{Tools for supersymmetry},'' {\em Ann. U. Craiova Phys.}
  {\bfseries 9} no.~I, (1999) 1--48,
\href{http://arxiv.org/abs/hep-th/9910030}{{\ttfamily arXiv:hep-th/9910030
  [hep-th]}}.

\end{thebibliography}\endgroup
\end{document}